# Imaging electron flow and quantum dot formation in MoS$_2$ nanostructures

*Sagar Bhandari[†], Ke Wang[†], Kenji Watanabe[ʝ], Takashi Taniguchi[ʝ],*

*Philip Kim[†], and Robert M. Westervelt[†,*]*

[†]School of Engineering and Applied Sciences and Department of Physics

Harvard University, Cambridge, MA 02138 USA

[ʝ]National Institute for Materials Science

1-1 Namiki, Tsukuba, 305-0044, Japan

**ABSTRACT:** Among newly discovered two-dimensional (2D) materials, semiconducting ultrathin sheets of MoS$_2$ show potential for nanoelectronics. However, the carrier mobility in MoS$_2$ is limited by scattering from surface impurities and the substrate. To probe the sources of scattering, we use a cooled scanning probe microscope (SPM) to image the flow of electrons in a MoS$_2$ Hall bar sample at 4.2 K. Capacitive coupling to the SPM tip changes the electron density below and scatters electrons flowing nearby; an image of flow can be obtained by measuring the change in resistance between two contacts as the tip is raster scanned across the sample. We present images of current flow through a large contact that decay exponentially away from the sample edge. In addition, the images show the characteristic "bullseye" pattern of Coulomb blockade conductance rings around a quantum dot as the density is depleted with a back gate. We estimate the size and position of these quantum dots using a capacitive model.

Ultrathin sheets of MoS$_2$ that are only a few atoms thick display unique electronic properties including a thickness- and strain-dependent bandstructure, valley Hall effects and spin-valley physics [1-3]. For graphene, covering both sides of a graphene sheet with layers of hexagonal boron nitride (hBN) greatly enhances the carrier mobility resulting in ballistic transport [4]. However, the measured mobility in an hBN encapsulated MoS$_2$ device is still limited to moderate values (500-2000 cm$^2$V$^{-1}$s$^{-1}$) [5-10]. Studies in MoS$_2$ [5-10] have shown that scattering from lattice defects, charged impurities, and substrate adsorbates lowers the mobility. Direct imaging of electron motion in MoS$_2$ devices can give us vital information about scattering, helping us to develop better devices. In previous research, we used a cooled scanning probe microscope (SPM) to image electron motion through a two-dimensional electron gas (2DEG) in a GaAs/AlGaAs heterostructure [11-14] and in graphene [15, 16]. In this approach, tip is raster scanned above the dot while the dot conductance is measured. We have also used our cooled SPM to image quantum dots formed in a GaAs 2DEG [17] and an InAs/InP nanowire [18] by using the tip as a scanning gate to tune the number of electrons on the dot, creating rings of high conductance about the dot corresponding to Coulomb-blockade conductance peaks [17, 18]. We have adapted both techniques to image electron flow and characterize quantum dots in a MoS$_2$ device.

In this paper, we present images of electron flow in a three-quantum-layer MoS$_2$ device at 4.2 K by using the tip to partially block the current. The device is a hBN-MoS$_2$-hBN sandwich patterned into a hall bar geometry (Fig. 1a). The tip is held 10 nm above the sample surface, creating an image charge in the MoS$_2$ that redirects electrons flowing nearby. When the tip is above the large contact (Fig. 1a), the tip partially blocks the current. The blocking action decays exponentially as the tip is moved away from the contact into the sample, as one might expect for a diffusive mean free path. As the carrier density is reduced toward the charge neutral point, we find that quantum dots are created in the small side contact (Fig. 1a): the SPM images show the characteristic "bullseye" pattern of Coulomb conductance peaks around each dot. Using the SPM images, we can locate each dot and determine its size.

**RESULTS AND DISCUSSION**

**Experimental Apparatus.** Figure 1a shows an optical image of the Hall-bar $MoS_2$ sample; the white ellipse and square indicate the regions of image scans. The Hall bar is patterned from a $hBN/MoS_2/hBN$ sandwich. It has dimensions 5.0x11.0 $\mu m^2$, with two narrow (1.0 $\mu$m) contacts along each side, separated by 3.0 $\mu$m, and large source and drain contacts (width 3.0 $\mu$m) at either end. The heavily doped Si substrate acts as a back-gate, covered by a 285 nm insulating layer of $SiO_2$. The back-gate capacitance is $C_G$ = 11.5 nF. The density $n$ can be tuned by applying an voltage $V_G$ between the backgate and the $MoS_2$ channel. The density $n$ is determined by hall measurements, using the side contacts.

To image electron motion using our cooled SPM, a voltage $V_s$ is applied between a side contact and the grounded source of the device. The at each tip position, the sample resistance $R = V_s/I_s$ is measured by the current $I_s$. To probe electron flow (Fig. 1b) the work function on the tip creates an image charge inside the $MoS_2$ channel with a corresponding potential (Fig. 1b) that scatters electrons away from their original trajectories, producing a change $\Delta R$ in the resistance. An image of electron flow is created by displaying $\Delta R$ as the tip is raster scanned above the sample at a constant height $h$. By reducing the electron density $n$ in the $MoS_2$ layer toward the charge neutral point, we observe how the electron flow changes.

**Images of Electron Flow.** Figure 2a shows an SPM image of electron flow through the wide contact for electron density $n = 1.3 \times 10^{12}$ cm$^2$ at temperature $T = 4.2$ K. The image area (white ellipse) is centered on the contact center and the edge of the MoS$_2$ channel outside the contact. Solid white lines in the image show the contact locations. The SPM image shows a peak in $\Delta R$ at the center that decays outward, indicating that the SPM tip partially blocks the current. A similar pattern is seen in the narrow side contacts. Figure 2b shows a semi-log plot of resistance change $\Delta R$ along a line through the middle of the partially blocked region of Fig. 2a *vs.* distance $d\ell$ into the sample. As shown, $\Delta R$ drops exponentially into the device with a characteristic length $L = 250$ nm.

It is interesting to compare the exponential decay in the image of Fig. 2b with the mean free path $\ell$ of electrons moving through a 2D electron gas. The resistivity is $\rho = (h/e^2)/k_F\ell$, where $e$ is the electron charge, $h$ is Planck's constant, $k_F$ is the Fermi wavevector, and $\ell$ is the mean free path. The Fermi wavevector is $k_F = (2\pi n)^{1/2}$, where $n$ is the electron density. Inserting into the conductivity we find:

$$\ell = \left(\frac{h}{e^2}\right)\frac{1}{\rho(2\pi n)^{1/2}}. \tag{1}$$

The sheet resistance $\rho$ is measured by a four probe Hall measurement with the current through the two wide contacts and the voltage measured between the two narrow contacts. For $n = 1.3 \times 10^{12}$ cm$^{-2}$, $\rho = 2{,}250$ $\Omega$, and the mean free path $\ell = 40$ nm from Eq. 1, well below the measured decay length $L = 250$ nm. The discrepancy between $\ell$ and $L$ may occur, because the density and mobility inside the contact are different from the interior of the device. The band structure of ultrathin MoS$_2$ sheets changes with thickness [1, 20, 21], the energy bands are sensitive to strain [22, 23] near the contacts, and the charge density can change near the edge [24].

Figure 3 shows SPM images of electron flow in the wide contact as the electron density $n$ is increased by tuning the backgate voltage from $V_G$ = 23.8 to 26.0 V. The images show no flow at the lowest density ($V_G$ = 23.8 V), because very few electrons are moving through the channel. As electrons are added ($V_G$ = 24.2 to 24.4 V) flow starts to appear in the images as electrons are blocked by the tip. The SPM images shows a maximum resistance change at $V_G$ = 24.8 V, then the flow starts to fade and eventually disappears at $V_G$ = 25.8 V. The maximum occurs, because the work function of the SPM tip creates a fixed image charge in the $MoS_2$ sheet. As the electron density increases from low values the image of flow becomes stronger, but the image fades away at higher densities, because the fixed image charge associated with the tip is a less effective scatterer.

**Quantum Dots.** As the electron gas inside the $MoS_2$ device is depleted to values near the charge neutral point, the SPM images reveal the presence of quantum dots associated with pools of electrons at low points in the background potential. Figure 4a shows an image of $\Delta G$ taken inside the narrow contact at the upper left side (Fig. 1a). A clear bullseye pattern of Coulomb blockade conductance peaks circle the location of a quantum dot; the tip is acting as a movable gate, and the number of electrons on the dot changes by one as the tip moves from one ring to the next. As the electron density is increased in Fig. 4b, a second quantum dot appears. Similar images of quantum dots were recorded previously for dots formed by top gates in a GaAs/AlGaAs heterostructure [17] and for an InAs dot formed in a InAs/InP nanowire [18].

Using a simple circuit model (Fig. 4c) we derive an expression to measure the radius of the quantum dot from the SPM images such as those shown in Fig. 4. The circuit includes the small tip-to-dot capacitance $C_{td}$ and the large backgate-to-dot capacitance $C_{gd}$ associated with the heavily doped Si substrate. The dot potential is $V_{dot}$, the backgate potential is $V_G$ and the tip potential is $V_{tip}$. Using a standard model, the conical tip is modeled by two conducting spheres at the same potential: a small sphere with the tip radius $a_{tip}$ and a much larger sphere representing the top of the cone. When the tip is scanned across the sample, the distance is larger than the tip diameter, but generally small compared with the large sphere radius; the tip motion provides the contrast, while the top of the cone provides a background level. The tip-to-dot capacitance is given by:

$$C_{td} = \frac{4\pi\epsilon_o a_{dot} a_{tip}}{r_{td}} \tag{2}$$

where $a_{dot}$ is the dot radius, $a_{tip}$ is the tip radius, and $r_{td}$ is the distance between the tip and the dot. Similarly, the backgate-to-dot capacitance is given by:

$$C_{gd} = \frac{4\pi\epsilon_o a_{dot}^2}{2d} \tag{3}$$

From the circuit model in Fig. 4(c), the dot charge $q_{dot}$ is:

$$q_{dot} = (C_{td}V_{tip} + C_{gd}V_G) \tag{4}$$

We apply two methods to induce a change $\Delta q_{dot}$ in the dot charge. Method #1 involves changing the tip position by $\Delta r_{td}$ to induce a change in dot charge $\Delta q_{dot}$ while keeping the backgate voltage $V_G$ fixed. For this case

$$\Delta q_{dot} = \frac{dq_{dot}}{dr_{td}} \Delta r_{td} = \frac{dC_{td}}{dr_{td}} \Delta r_{td} \tag{5}$$

Therefore, charge induced in the dot by a change in tip position $\Delta r_{td}$ becomes

$$\Delta q_{dot} = (V_{tip} + V_G) \frac{4\pi\epsilon_o a_{dot} a_{tip}}{r_{td}^2} (\Delta r_{td}) \tag{6}$$

Method #2 involves inducing $\Delta q_{dot}$ by a change in the backgate voltage $V_G$ keeping the tip position fixed. For this method

$$\Delta q_{dot} = \frac{dq_{dot}}{dV_G} \Delta V_G = C_{gd} \Delta V_G \qquad (7)$$

**Estimation of Dot Size.** Cooled SPM images of bullseye pattern of conductance rings are shown in Fig. 4(a) and 4(b). In these images, the backgate voltage is kept fixed at $V_G = 4.80$ V and $V_G = 5.29$ V respectively. In Fig. 4(a), a single quantum dot is located at the center of the bullseye. Each conductance ring corresponds to an electron being added to the dot $\Delta q_{dot} = e$ by changing the tip-to-dot capacitance *via* tip motion. Using Method #1, from the spacing between these rings and their distance from the center, we can compute the size and position of the dot. Figure 4(d) A plot of the ring spacing $\Delta r_{td}$ *vs.* $r_{td}^2$ shows a linear dependence which agrees well with Eq. 7. The slope determines the dot radius $a_{dot} = 180$ nm, using $V_{tip} = -1.00$ V and $a_{tip} = 10$ nm.

For Method #2, we keep the tip position fixed and change the backgate voltage $V_G$. Figure 5(a) shows a series of SPM images of $\Delta G$ in the same location as Fig. 4 for backgate voltages ranging from (a) $V_G = 4.80$ V to (h) $V_G = 5.29$ V. An additional quantum dot appears as the density is increased. To measure the effect of changing $V_G$, we pick a fixed tip position $X = -0.5$ µm, $Y = 0.5$ µm. Figure 5i plots $\Delta G$ at this tip position *vs.* $V_G$. Figure 5i shows five peaks, and each peak corresponds to the addition of one electron charge $e$ to the quantum dot. To get the peak spacing, the peak position in $V_G$ vs. the peak number is plotted. The slope of this line gives the average peak spacing $\Delta V_G = 50$ mV. By putting the average peak spacing in $V_G$ into Eq. 7, we obtain the quantum dot radius $a_{dot} = 150$ nm, in good agreement with the dot radius found by Method #1.

**CONCLUSION**

The unique properties of MoS$_2$ [1-7] open the way for new electronic and photonic devices, as well as an opportunity to probe the physics of atomic layer systems. These properties include thickness/strain dependent band structure [1, 20, 21], valley Hall effects [2] and spin-valley physics [3]. The measured carrier mobility of $MoS_2$ devices is low despite encasing the $MoS_2$ sheet between two hBN layers. In our imaging experiment, we use a cooled SPM to image electron motion in a few atomic layer $MoS_2$ device. The scanning tip partially blocks the electron flow close to the contacts, a region where the contact meets the channel, thereby increasing the resistance of the channel. Displaying the resistance change *ΔR vs.* tip position produces a map of electron flow. The resistance change *ΔR*, imaged in large contact at the bottom of Fig. 1a, drops exponentially as the tip is moved into the sample, with a decay length $L = 250$ nm. The mean free path $ℓ = 40$ nm estimated from the sample resistivity $\rho$ and density $n$ is smaller. This difference could be due to a variation in density and band structure of $MoS_2$ near the edge and contacts.

At low electron density near the charge neural point, quantum dots form in the narrow side contacts. We observe the characteristic bullseye pattern of Coulomb conductance peaks from two quantum dots formed in the narrow contact at the upper left of Fig. 1a. Using a capacitive model, we estimate the dot radius using two methods to be $a_{dot} = 180$ nm and $a_{dot} = 150$ nm, in good agreement. The quantum dots are presumably formed by pools of electrons at minima in the background potential.

This paper demonstrates how a cooled SPM can image electrons flow inside a $MoS_2$ Hall bar. A similar technique could be used to map electron flow through a wide variety of devices made from $MoS_2$ as well as other semiconducting transition metal dichalcogenides, such as $WSe_2$. The cooled SPM can also image the presence of quantum dots created at low densities by roughness in the background potential, giving their location and radius, using our previously developed technique [17,18].

**METHODS**

**Device Fabrication.** Using a dry transfer technique, we assembled a van der Waals (vdW) heterostructure consisting of a few quantum layer $MoS_2$ sheet encased by two insulating hBN layers. The assembly is then transferred to a heavily doped silicon wafer covered with a $SiO_2$ layer that is 285 nm thick with pre-deposited gates: 1 nm Cr layer followed by a 7 nm layer of Pd-Au alloy (40% Pd and 60% Au, by wt.) covering most of the graphene regions. The device is subsequently vacuum-annealed at 350 ºC to reduce structural inhomogeneity. Finally, the Hall bar geometry is defined by reactive ion etching and 1D edge contacts to each graphene layers are fabricated with Cr/Pd/Au (1.5 nm/5 nm/120 nm) metal deposition.

**Cooled Scanning Probe Microscope.** We use a home-built cooled scanning probe microscope (SPM) to image the motion of electrons in our sample. The microscope assembly consists of a head assembly where the tip is attached and a cage assembly enclosing the piezotube translator that scans a sample fixed on top in the *X*, *Y* and *Z* directions. Scans are performed by actuating the piezotube with home-built electronics including an *X-Y* position controller for scanning, and a feedback *Z* controller for topological scans of the sample surface [13, 16]. The microscope assembly is placed in an insert inside a liquid He Dewar; the insert is filled with 3.0 mbar of He exchange gas to cool the sample and SPM. For the transport measurements, standard lock-in amplifiers are used. For the scanning gate measurements, an SPM tip of 10 nm radius was held at a fixed height 10 nm above the BN surface, which is approximately 50 nm above $MoS_2$ layer. To create an image, the resistance was displayed while the tip was raster scanned across the sample.

**AUTHOR INFORMATION**

**Corresponding Author**

Robert M. Westervelt

*Email: westervelt@seas.harvard.edu

**Notes**

The authors declare no competing financial interest.

**ACKNOWLEDGEMENTS**

The SPM imaging experiments and the ray-tracing simulations were supported by the U.S. DOE Office of Basic Energy Sciences, Materials Sciences and Engineering Division, under grant DE-FG02-07ER46422. The $MoS_2$ sample fabrication was supported by Air Force Office of Scientific Research contract FA9550-14-1-0268 and Army Research Office contract W911NF-14-1-0247. Growth of hexagonal boron nitride crystals was supported by the Elemental Strategy Initiative conducted by the MEXT, Japan and a Grant-in-Aid for Scientific Research on Innovative Areas No. 2506 "Science of Atomic Layers" from JSPS. Nanofabrication was performed in the Center for Nanoscale Systems (CNS) at Harvard University, a member of the National Nanotechnology Coordinated Infrastructure Network (NNCI), which is supported by the National Science Foundation under NSF award ECCS-1541959.

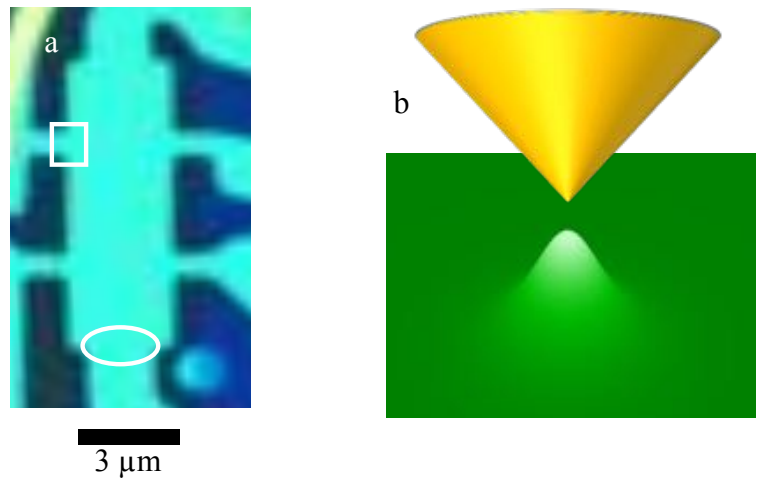

**Figure 1**: (a) Optical image of the hBN/MoS$_2$/hBN device patterned into a Hall bar geometry. The white outlines indicate the regions where SPM imaging experiments were performed. (b) Schematic diagram showing the potential inside the MoS$_2$ layer created by the SPM tip. This potential deflects electrons flowing nearby creating a change $\Delta R$ in the device resistance. An image is formed by displaying $\Delta R$ vs. tip position as the tip is raster scanned across the sample.

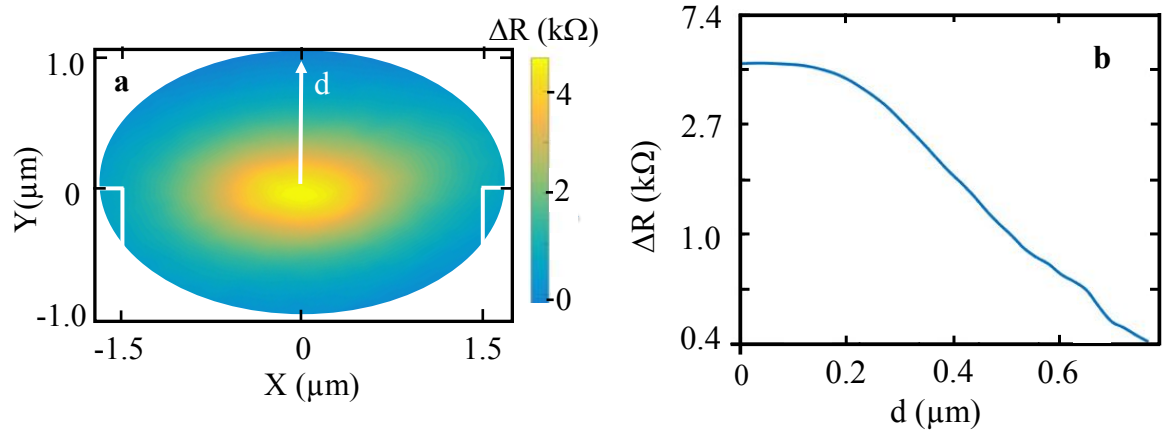

**Figure 2**: (a) Image of resistance change $\Delta R$ vs. tip position in the wide contact (b) Semi-log plot of resistance change $\Delta R$ measured at $X = 0$ μm vs. distance $d$ into the device from the sample edge; $\Delta R$ decays exponentially into the device.

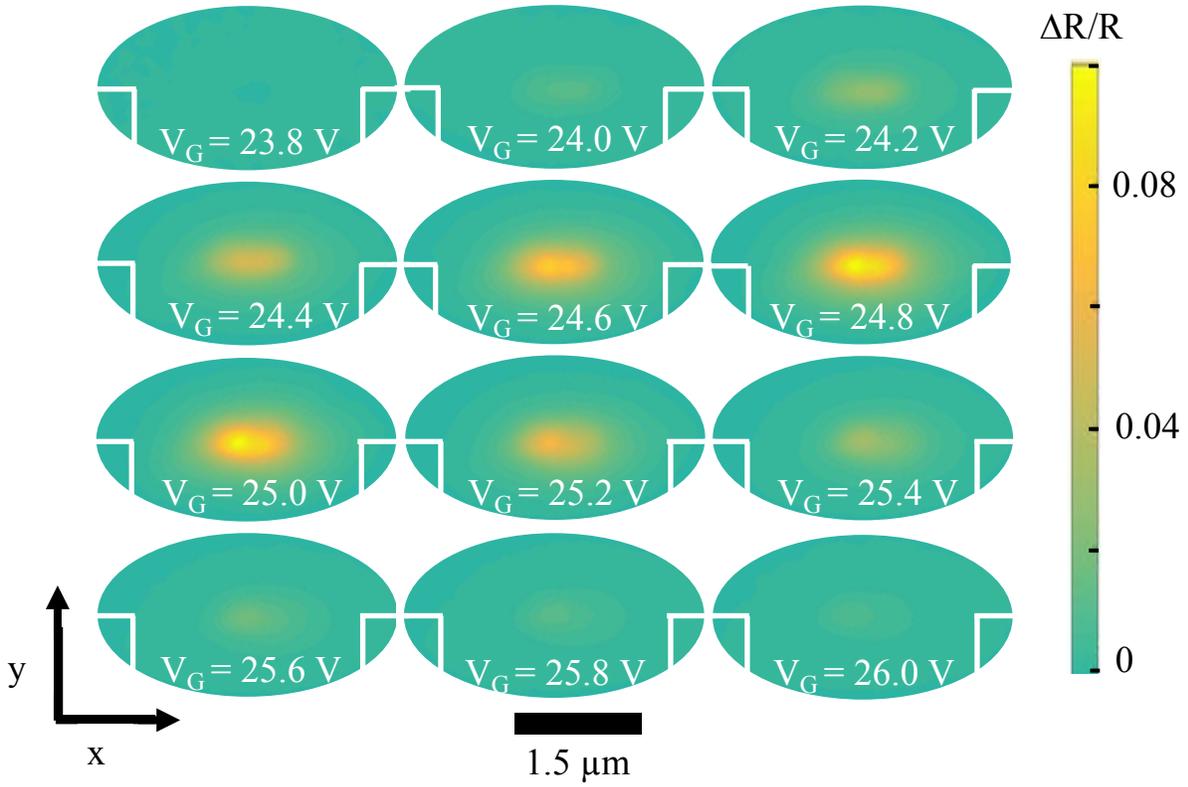

**Figure 3**: Tiled plot of the tip-induced change $\Delta R/R$ vs. tip position at the end of the wide contact with more positive backgate voltage $V_G$. The images show that $\Delta R/R$ initially increases with electron density, reaches a maximum, and drops as the carrier density is increased further.

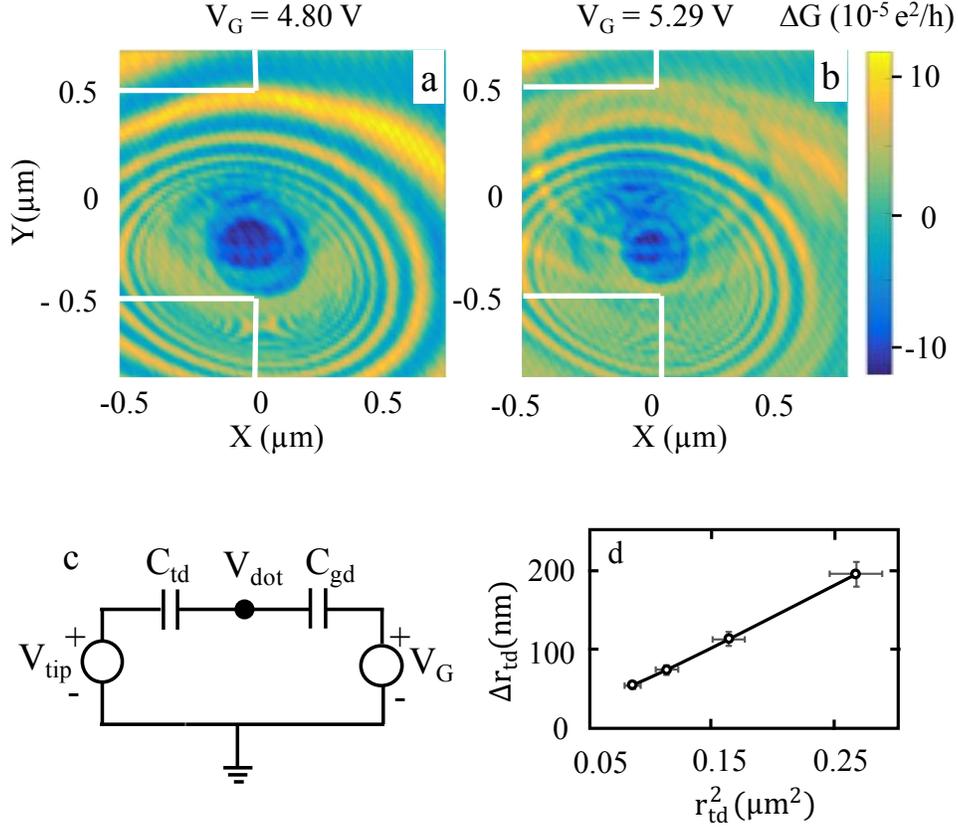

**Figure 4**: (a) Display of conductance change $\Delta G$ vs. tip position at the narrow contact (see Fig. 1) for $V_G = 4.80$ V, when the device is nearly depleted. The bullseye pattern of concentric rings are Coulomb blockade conductance peaks associated with a quantum dot at the center. (b) As the density is increased for $V_G = 5.29$ V an additional quantum dot appears at a different location. (c) Schematic circuit model of a quantum dot, showing the tip-to-dot capacitance $C_{td}$ and the backgate-to-dot capacitance $C_{gd}$ (see Eq. 4). (d) Spacing $\Delta r_{td}$ between conductance rings vs. radial distance $r_{td}$ between the tip and bullseye center. The measured dot radius is $a_{dot} = 180$ nm.

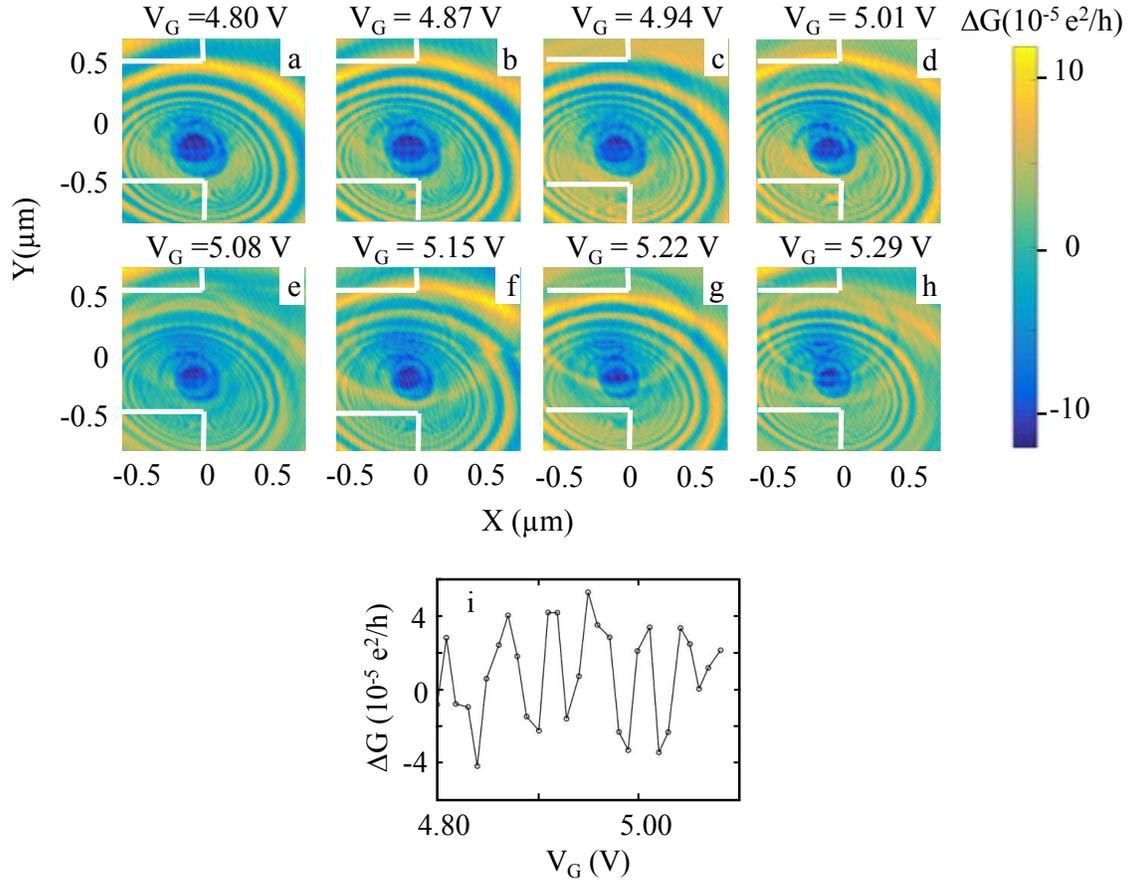

**Figure 5:** Images that display $\Delta G$ vs. tip position in the narrow contact at the same locations as Fig. 4 for a series of backgate voltages indicated on the figure, ranging from (a) $V_G = 4.80$ V to (h) $V_G = 5.29$ V. The bullseye pattern of Coulomb conductance peaks in (a) shows the existence of a quantum dot. A second dot is created as $V_G$ is increased. (i) Plot of the conductance change $\Delta G$ from the series of images at tip position $X = -0.5$ µm, $Y = 0.5$ µm vs. $V_G$. To get the peak spacing, the peak position in $V_G$ vs. peak number is plotted and slope of this line gives average peak spacing $\Delta V_G = 50$ mV. Using the expression for charge induced in the quantum dot as $V_G$ is varied, the measured dot radius is $a_{dot} = 150$ nm in good agreement with Fig. 4.